\documentclass{iau}
\usepackage{graphicx}

\title[IMBHs: understanding the data first] 
{Searching for intermediate mass black holes: understanding the data first}

\author[Paolo Bianchini \etal\ ]   
{
Paolo Bianchini, Mark Norris, Glenn van de Ven, \and Eva Schinnerer}

\affiliation{Max-Planck Institute for Astronomy, K\"{o}nigstuhl 17, 69117 Heidelberg, Germany \\email: {\tt bianchini@mpia.de}}

\pubyear{2015}
\volume{312}  
\pagerange{xx--xx}
\jname{Star Clusters and Black Holes in Galaxies across Cosmic Time}
\editors{}
\begin{document}

\maketitle

\begin{abstract}
The detection of intermediate mass black holes (IMBHs) in globular clusters has been hotly debated, with different observational methods delivering different outcomes for the same object.
In order to understand these discrepancies, we construct detailed mock integral field spectroscopy (IFU) observations of globular clusters, starting from realistic Monte Carlo cluster simulations. The output is a data cube of spectra in a given field-of-view that can be analyzed in the same manner as real observations and compared to other (resolved) kinematic measurement methods.
We show that the main discrepancies arise because the luminosity-weighted IFU observations can be strongly biased by the presence of a few bright stars that introduce a scatter in velocity dispersion measurements of several km s$^{-1}$. We show that this intrinsic scatter can prevent a sound assessment of the central kinematics, and therefore should be fully taken into account to correctly interpret the signature of an IMBH.

\end{abstract}

\firstsection 
\section{Introduction}
The existence of intermediate mass black holes (IMBHs) with masses between those of stellar black holes (M$_\bullet<100$ M$_\odot$)
and those of super massive black holes (SMBH, M$_\bullet>$ 10$^5$ M$_\odot$) has been postulated. The extrapolation of the $M_\bullet-\sigma$ relation for galaxies, linking the mass of the central back hole to the velocity dispersion of the host stellar system (\cite{Ferrarese2000,Magorrian1998}), suggests that globular clusters (GCs) represent the ideal environments to find central IMBHs with masses ranging from $10^3-10^4$~M$_\odot$. However, their detection has proven to be difficult, with contradictory results on their presence in local group GCs (e.g., \cite{Gebhardt2000,vandenBosch2006,Noyola2010,vanderMarel2010,Luetzgendorf2013,Lanzoni2013,denBrok2014}). 

The kinematic observations suggesting detections of IMBHs are primarily based on the search for a rise of the central velocity dispersion (e.g. \cite{Bahcall1976,Luetzgendorf2013}). This method is very challenging, since it requires both high spatial resolution, to resolve the very crowded central region of GCs (few central arcseconds), and very precise velocity measurements with accuracy $\sim1$ km s$^{-1}$.

Two distinct observational strategies are employed for the kinematic detection of IMBHs: 
1) measurements of velocities of resolved individual stars (line-of-sight velocities or proper motions), 2) unresolved kinematic measurements with integral field unit (IFU) spectroscopy, from line  broadening of integrated spectra.
These complementary methods can give significantly different observational outcomes when applied to the same object, 
making the detection of IMBHs highly ambiguous. In particular, integrated light spectroscopy seems to measure rising central velocity dispersions, favoring the presence of IMBHs (see for example, \cite{Noyola2010} for $\omega$ Cen, or \cite{Luetzgendorf2011} for NGC 6388), while resolved stellar kinematics do not confirm the presence of this signature (see \cite{vanderMarel2010} for 
proper motion measurements of $\omega$ Cen, and \cite{Lanzoni2013} for discrete line-of-sight measurements in NGC 6388).

Our goal is to understand the systematic differences between the different observational methods, before undertaking any interpretation of the kinematic signatures connected to the presence of IMBHs. In particular, we wish to understand the biases that arise from applying IFU spectroscopy to systems with a (partially) resolved stellar population, like Galactic GCs.
In order to do so, we develop a procedure to create detailed mock IFU observations of the center of GCs starting from realistic Monte Carlo cluster simulations. The output of our procedure is a data cube with spectra and luminosity information for every spaxel in a selected field-of-view, that will be analyzed in the same manner as real IFU observations to build mock velocity dispersion profiles in the central region of a GC.

\section{Constructing a mock IFU observation}

\begin{figure}[t]
\begin{center}
\includegraphics[width=0.40\textwidth]{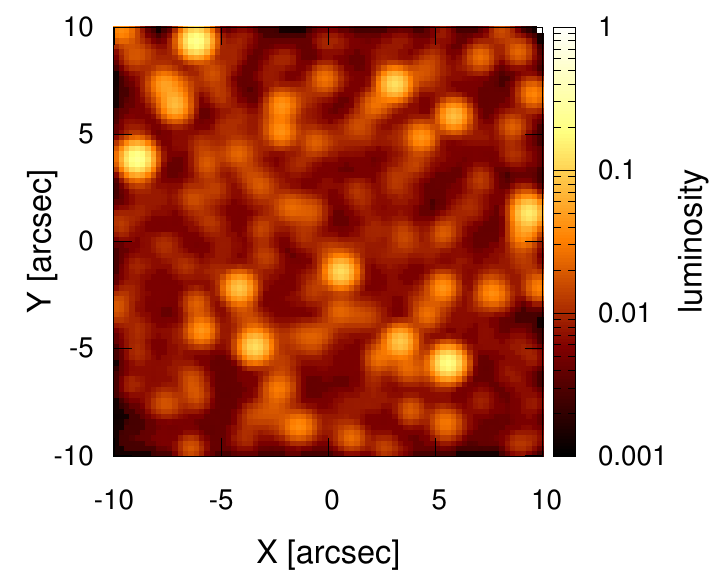}\,
\includegraphics[width=0.58\textwidth]{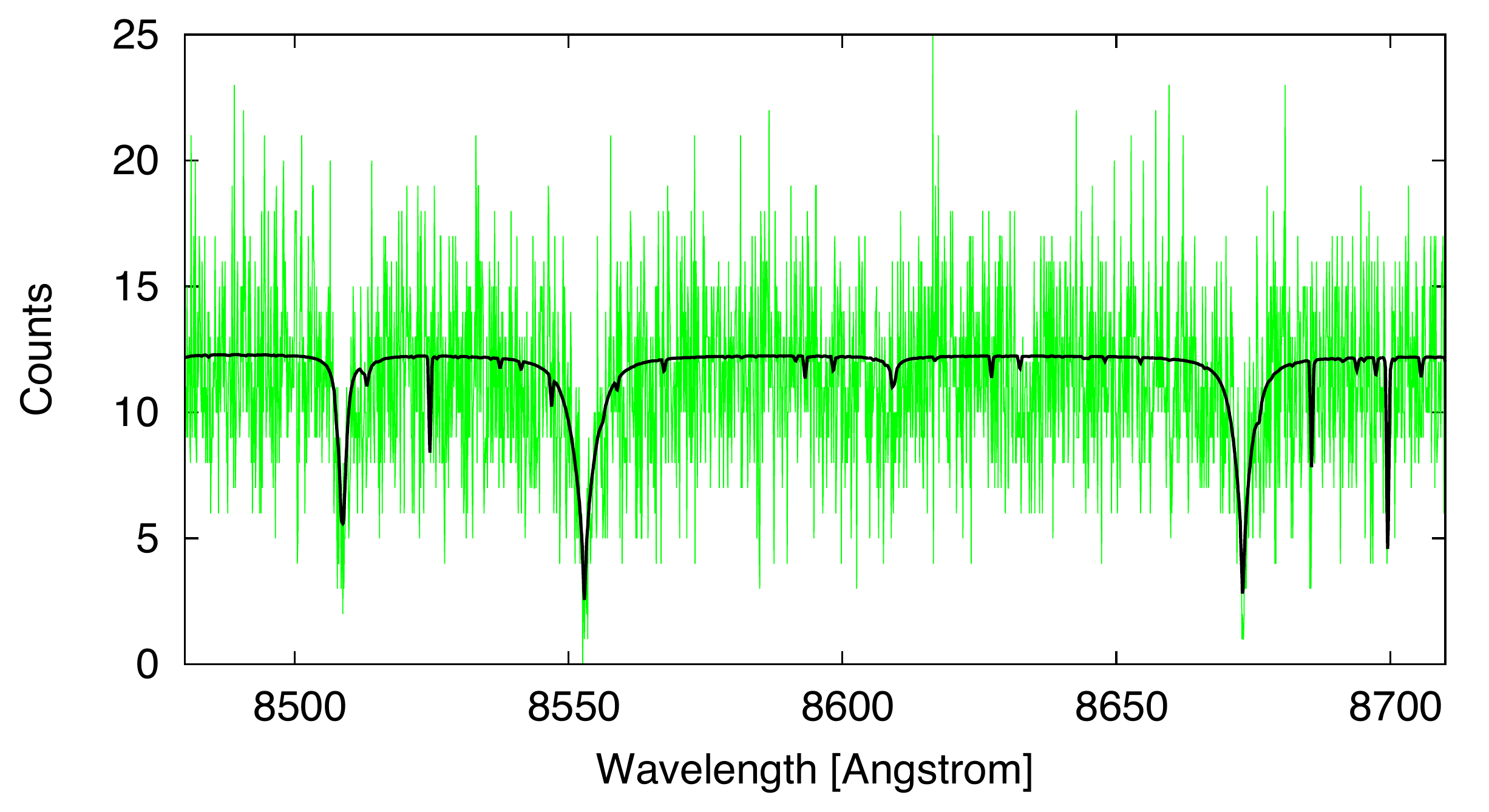}
\caption{\textit{Left panel:} Luminosity map of the central $20\times20$ arcsec$^2$ region of our simulated globular cluster placed at 10 kpc and observed with a seeing of 1 arcsec and an average signal-to-noise per {\AA} of $S/N\approx10$. \textit{Right panel:} typical spectrum of a spaxel, obtained summing all the Doppler-shifted spectra falling in the spaxel, properly weighted by their PSF. The black line indicates the spectrum without noise, while the green line indicates the case of an observation with signal to noise per {\AA} of $S/N\approx10$.}
\label{fig1}
\end{center}
\end{figure}

The starting point of our work are Monte Carlo cluster simulations, developed by \cite{Downing2010}, providing a realistic description of a typical GC with initial number of particles of $2\times10^6$ drawn from a \cite{Plummer1911} model, a \cite{Kroupa2001} initial mass function, 10\% primordial binary fraction, and metallicity [Fe/H]=-1.3. The simulations have no central IMBH and no internal rotation (note however that internal rotation is observed in several GCs, e.g. \cite{Bianchini2013, Fabricius2014, Kacharov2014}). At 13 Gyr, the simulation is characterized by a total mass of $M\simeq6.7\times10^5$ $M_\odot$ and a projected half light radius of $R_h\simeq2.8$ pc. We place the simulated GC at 10 kpc from the observer with a global systemic line-of-sight velocity of 300 km s$^{-1}$, to match the typical properties of a Galactic GC.

We associate to each star (characterized by effective temperature $T_\mathrm{eff}$, mass $M_\star$, luminosity $L_\star$, metallicity $Z$) a low-resolution broad-wavelength stellar spectrum, using GALEV evolutionary synthesis model (\cite{Kotulla2009}). Then we associate a high-resolution spectrum in the wavelength range that will be used in our mock observations (calcium triplet, 8400-8800 {\AA}) using the MARCS synthetic stellar library (\cite{Gustafsson2008}) providing a resolving power of $R=\lambda/\Delta\lambda=20\,000$, enough to measure the internal kinematics of GCs (typical velocity dispersions of 10 km s$^{-1}$).
Finally, we Doppler-shift the spectra using the line-of-sight velocity from our cluster simulation. 

We next define the observational setup of the simulated IFU instrument, selecting a $20\times20$ arcsec$^2$ field-of-view and a spaxel scale of 0.25 arcsec.
After convolving each star with a Gaussian PSF, we sum the Doppler-shifted spectra of all the stars falling in each spaxel properly weighted by their PSF. In this way we have a spectrum and the 
corresponding luminosity information for each spaxel. Finally, we add Poisson noise to the final spectra in order to match the desired signal-to-noise ratio $S/N$. In Fig. \ref{fig1} we show the final product of our mock IFU observation, consisting of a luminosity map, with seeing of 1 arcsec and average signal-to-noise per {\AA} of $S/N=10$ (left panel), and the typical spectrum of a spaxel (both with and without noise; right panel).

\section{Analyzing the kinematics}

\begin{figure}[t]
\begin{center}
 \includegraphics[width=1\textwidth]{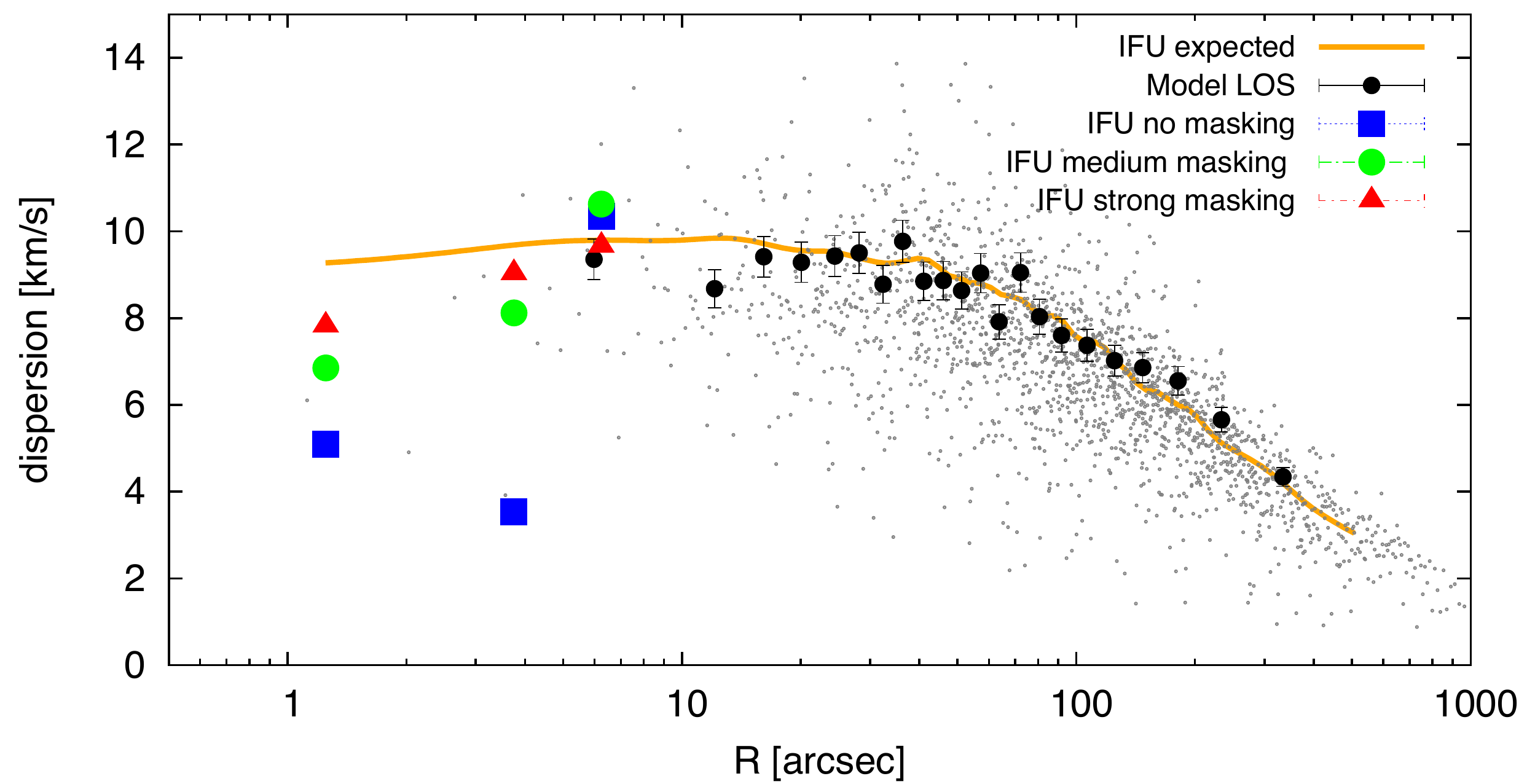} 
 \caption{Velocity dispersion profiles for our mock IFU observation and our model GC. Both a discrete velocity dispersion profile based on giants stars (filled circles) and a luminosity-weighted profile (grey dots) constructed directly from our model are shown. The solid line is the profile expected for the typical kinematical tracer of an IFU observation (\cite{Bianchini2015}). The luminosity-weighted profile shows a high scatter due the luminosity differences between stars. The profiles in the central region are extracted from our mock IFU observation (blue squares), and recalculated after masking the brightest spaxels (green circles and red triangles).}
   \label{fig2}
\end{center}
\end{figure}

After producing the data cube of our IFU observation, we can construct the one-dimensional velocity dispersion profiles. We divide our field of view in annular bins and sum all the spectra in each bin. With the penalized pixel-fitting (pPXF) program of \cite{Cappellari2004} we measure the velocity dispersion from the broadening of the lines of the summed spectra. The measured velocity dispersion profile of the central region can then be directly compared to the dispersion profiles expected from the model.

We construct three different velocity dispersion profiles from the model. The first is a typical line-of-sight velocity dispersion profile, constructed using the resolved kinematics of only the bright giant stars in our simulation. The second is a luminosity-weighted velocity dispersion profile, constructed taking into consideration all stars in our simulation and their corresponding luminosity. The third is the profile expected for the typical kinematical tracer of an IFU observation (\cite{Bianchini2015}). Fig. \ref{fig2} shows that the luminosity-weighted profile is characterized by a large scatter due to the stochasticity introduced by the luminosity differences between stars.

Since an IFU observation gives intrinsically luminosity-weighted kinematic measurements, it is already evident that it will suffer from a stochasticity effects driven by the presence of a few bright stars that can completely dominate certain spaxels. A procedure often employed to minimize the stochasticity consists in excluding from the kinematic analysis the spaxels that are found contaminated by a bright star. A simple strategy consists in masking the brightest spaxels in our IFU field-of-view and then recompute the velocity dispersion profile. We present in Fig. \ref{fig2} the profiles obtained in the IFU field-of-view without masking, and after applying a medium and a strong masking (eliminating respectively 10\% and 30\% of the brightest spaxels). The figure shows that masking can significantly change the observed central velocity dispersion.

\section{Conclusion}

We have presented a procedure to simulate IFU observations of Galactic GCs. Our procedure allows to obtain a data cube of spectra in a selected field-of-view starting from any GC simulation. We used a mock IFU observation of a realistic 10$^6$ particle Monte Carlo cluster simulation to investigate what are the effects that can bias the kinematic measurements. Our analysis shows that IFU kinematic measurements can be strongly biased by the presence of a few bright stars, that can dominate certain spaxels. Moreover IFU kinematics gives a luminosity-weighted information, and therefore can in principle give different outcomes from what obtained from resolved line-of-sight kinematics. These stochasticity effects can prevent to obtain sound measurements of the central velocity dispersion of GCs and therefore must be carefully investigated before interpreting any signatures of the presence of IMBHs.
We will use our procedure to study in detail the stochasticity connected to luminosity-weighted kinematics and to determine an efficient masking technique to recover unbiased results. Moreover, we will compare mock observations of models with and without IMBHs to help understanding if the presence of an IMBH can be definitively recovered using common dynamical modeling techniques.

\end{document}